\documentclass[pra,twocolumn,nofootinbib,floatfix,10pt]{revtex4-2}

\usepackage{amsmath}
\usepackage{amssymb}
\usepackage{wasysym}
\usepackage{graphicx}
\usepackage{color,soul}
\usepackage{physics}
\usepackage{siunitx}
\usepackage{dsfont}
\usepackage{float}
\usepackage{xcolor}
\usepackage[english]{babel}
\usepackage{blindtext}
\usepackage[english,nomargin,inline,marginclue,draft]{fixme}
\pdfpageheight\paperheight
\pdfpagewidth\paperwidth

\usepackage[colorlinks,linkcolor=blue,anchorcolor=blue,citecolor=blue,urlcolor=blue]{hyperref}

\fxusetheme{colorsig}
\FXRegisterAuthor{cg}{acg}{CG}  
\FXRegisterAuthor{th}{ath}{\color{blue}TH}  
\FXRegisterAuthor{ib}{aib}{\color{red}IB} 
\FXRegisterAuthor{sh}{ash}{\color{cyan}SH} 
\FXRegisterAuthor{db}{adb}{\color{green}DB} 
\FXRegisterAuthor{ps}{aps}{PS}
\makeatletter
\renewcommand*\FXLayoutInline[3]{%
  {\@fxuseface{inline}\ignorespaces{\color{fx#1}[#3: #2]}}}
\makeatother

\long\def\symbolfootnote[#1]#2{\begingroup%
\def\thefootnote{\fnsymbol{footnote}}\footnotetext[#1]{#2}\endgroup}

\def\nobreakbefore{%
  \relax\ifvmode\else
    \ifhmode
      \ifdim\lastskip > 0pt\relax
        \unskip\nobreakspace
      \else 
        \nobreakspace
      \fi
    \fi
  \fi
}
\let\oldcite\cite
\renewcommand\cite{\nobreakbefore\oldcite}

\begin{document}
\title{All-optically tunable enantio-selectivity and chirality transfer}

\author{En-Ze Li$^{1,2,\textcolor{blue}{\star}}$}
\author{Ming-Xin Dong$^{1,2,\textcolor{blue}{\star},\textcolor{blue}{\S},\textcolor{blue}{\P}}$}
\author{Dong-Sheng Ding$^{1,2,\textcolor{blue}{\dagger}}$}
\author{Bao-Sen Shi$^{1,2,\textcolor{blue}{\ddagger}}$}
\author{Guang-Can Guo$^{1,2}$}
\author{Franco Nori$^{3,4}$}

\affiliation{$^1$Key Laboratory of Quantum Information, University of Science and Technology of China, Hefei, Anhui 230026, China.}
\affiliation{$^2$Synergetic Innovation Center of Quantum Information and Quantum Physics, University of Science and Technology of China, Hefei, Anhui 230026, China.}
\affiliation{$^3$Center for Quantum Computing, and Cluster for Pioneering Research, RIKEN, Wakoshi, Saitama 351-0198, Japan.}
\affiliation{$^4$Physics Department, University of Michigan, Ann Arbor, MI 48109-1040, USA.}

\date{\today}

\symbolfootnote[1]{E.Z.L and M.X.D contribute equally to this work.}
\symbolfootnote[5]{dongmx@ustc.edu.cn}
\symbolfootnote[2]{dds@ustc.edu.cn}
\symbolfootnote[3]{drshi@ustc.edu.cn}
\symbolfootnote[4]{present address: Laboratoire Kastler Brossel, Sorbonne Université, CNRS, ENS-Université PSL, Collège de France, 4 Place Jussieu, 75005 Paris, France.}

\maketitle

\textbf{Detecting and controlling the chirality of materials play an essential role in exploring nature, providing new avenues for material creation, discrimination, and manipulation.
In such tasks, chiral reagents are essential in defining or enhancing the chiral dichroism response. 
However, ignoring their influences on the symmetry of the medium hamper the ability to control and induce asymmetric synthesis.
Here, we propose a simple but versatile chirality transfer method for synthesizing and manipulating the chirality of medium. 
The proposed method induces the dispersion of light in a neutral atomic system, allowing to deterministically and tunably control the chirality transfer using a helical field.
First, we theoretically analyze the mechanism for this optically induced chirality transfer. 
Afterwards, we experimentally study the enantio-sensitive feature of the medium exposed to the auxiliary chiral field.
This result can be suppressed or enhanced in a deterministic enantio-selection, opening up an efficient way to manipulate asymmetric synthesis.
}

Chiral optics has attracted considerable scientific interest in life sciences \citep{kissick2011second,chen2009nanoparticle,evers2022theory,blackmond2019autocatalytic,guijarro2008origin,scully1999quantum,agarwal2012quantum,shreiner2022electrically},  the synthesis of chiral materials \citep{calavalle2022gate,nakashima2022self,shreiner2022electrically},  and sensitive detection and classification \citep{grier2003revolution,hendry2010ultrasensitive,mohammadi2018nanophotonic,tang2011enhanced,ye2021enhancing,ayuso2019synthetic,tkachenko2014optofluidic}. 
The enantiomeric medium is distinguishable when it interacts with chiral fields, and the chiroptical effects become quantitatively measurable in such enantio-sensitive process \citep{bliokh2011characterizing,tang2010optical,bliokh2013dual,bliokh2017optical,alpeggiani2018electromagnetic}.
In essence, the performance of the chiroptical effect depends on the complexity of the electromagnetic field distribution structure \citep{yang2009spectroscopy,barr2018investigating,bliokh2013dual,bliokh2014magnetoelectric,alpeggiani2018electromagnetic,vazquez2018optical}. 
Many efforts have been devoted to improving the chiral characterization model and proposing novel chiral manipulation schemes \citep{bliokh2013dual,bliokh2014magnetoelectric,alpeggiani2018electromagnetic,vazquez2018optical}. 
The latter helps to design and improve chiral self-assembly and asymmetric synthesis.

\begin{figure*}[t]
\includegraphics[width=0.8\paperwidth]{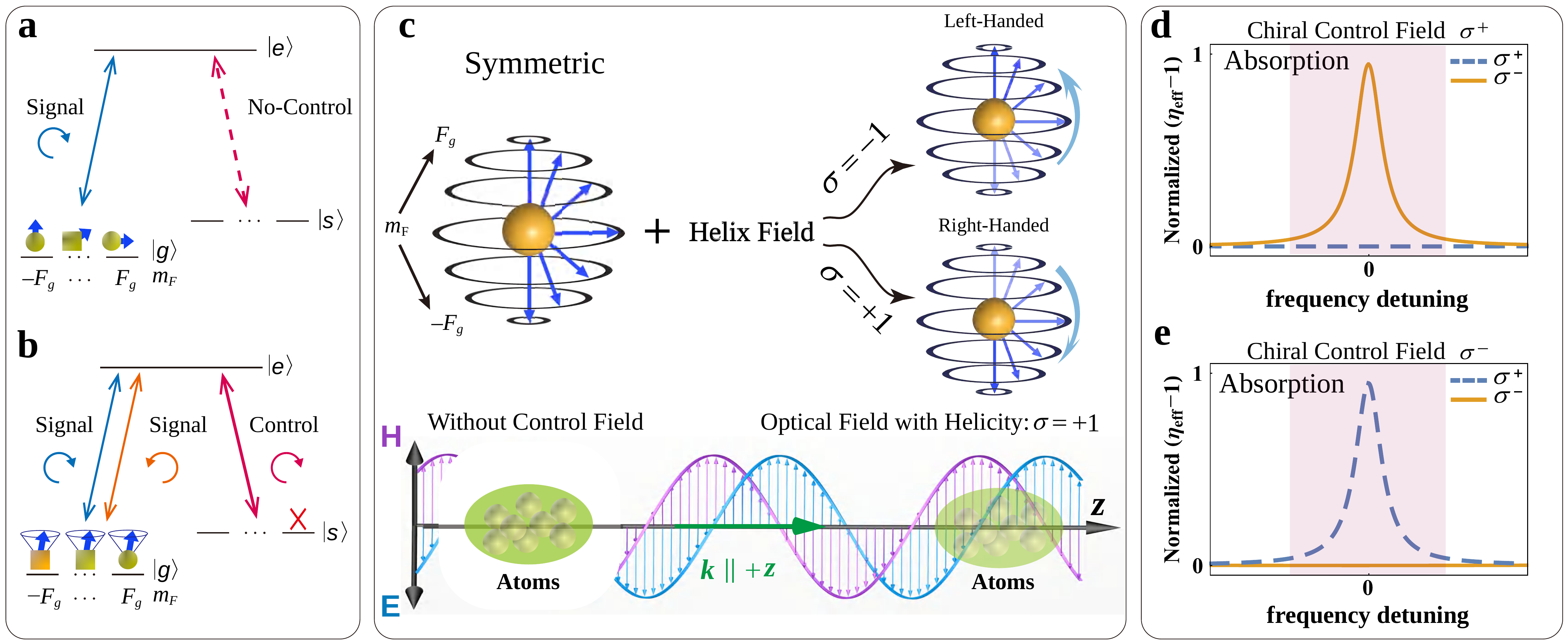}
\caption{\textbf{Schematic of the chirality transfer mechanism.} 
\textbf{a,b}, The energy diagram of the typical atomic medium, with degenerate ground Zeeman levels, $\left|g\right\rangle $ and $\left|s\right\rangle $, and the excited state $\left|e\right\rangle$ allows a double photon resonance transition. 
The control field couples $\left|s\right\rangle $ and $\left|e\right\rangle $, inducing the chirality-transfer process from the chiral field to the homogeneous medium.
This leads to a different response to the $\sigma^{+}$ and $\sigma^{-}$ polarized signals (coupling $\left|g\right\rangle $ and $\left|e\right\rangle $) due to the enantio-sensitive selection rules. 
\textbf{c}, Without an external chiral control field, the isotropic medium shows homogeneous absorptive loss and is insensitive to the enantiomeric signal fields. 
Applying an external chiral control field, the medium exhibits a chiral feature (enantio-sensitive absorption). 
In addition, the medium exhibits an inhomogeneous photon-enhanced chirality. 
\textbf{d,e}, Illustration of circular dichroism for $\sigma^{+}$ (blue dashed curve) and $\sigma^{-}$ (orange solid curve) polarization signal photons for the enantiomeric control fields, respectively. The enantiomeric transmission spectrum of the signal field is dominated by the chiral property of the control field.}
\label{Fig1}
\end{figure*}

Chirality transfer, a strategy of asymmetric synthesis, has gained much attention due to its deterministic enantio-selectivity, and ability to transfer the medium chirality within or between chirality elements \citep{hendry2012chiral,mohammadi2018nanophotonic}. 
Chirality transfer was widely explored in the area of magnetohydrodynamic turbulence \citep{moffatt1992helicity}, the production of chiral medicine \citep{farina2006asymmetric}, a single-mirror isomer in materials \citep{brown1989chemical}, the developments of chiral materials \citep{green1979asymmetric}, the purification of enantiopure samples \citep{ozturk2022origins,howard2022data,perez2018state}, etc. 
Unfortunately, it is difficult to dynamically control the chirality transfer, where the enantiomeric population difference in the enantio-selective excitation process is merely at a level of a few percent \citep{eibenberger2017enantiomer,perez2017coherent,perez2018state}. 
There is no efficient and fully controllable transfer paradigm, which limits the widespread use of chirality transfer strategies. 
Our experiment herein demonstrates the possibility of using the optical field as a polarization tool to promote the asymmetric synthesis of the medium.

We demonstrate a proof-of-concept experiment of a chirality transfer method with exceptional enantio-selectivity. 
This method relies solely on chiral electric-dipole transitions and works globally throughout the interaction region. 
The helical auxiliary field controls the macroscopic medium chirality at the single-photon level, maximizing it in one enantiomer while suppressing it in its mirror twin. 
On the contrary, if the helicity of the control light is flipped, the sign of the medium helicity will be accordingly reversed.

Note that such a method is based on a universal symmetry-breaking mechanism in the chiral optical interaction region, which exploits the helical control field to inherently break duality symmetry.
Furthermore, the chirality-transfer property is suitable for the case in the quantum regime. 
This might be helpful for the construction of chiral quantum devices.
The proposed approach is universal for a variety of physical systems, e.g., the detection and separation of enantiomeric excess \citep{perez2018state,fanood2015enantiomer}, and chirality transfer devices in quantum photonics \citep{sayrin2015nanophotonic,scheucher2016quantum,li2018quantum,li2020experimental,dong2021all}, which also provide an avenue for studying chiral and topological properties induced by the auxiliary field \citep{orlova2015creation,eismann2021transverse,ozawa2019topological,xia2018cavity,wang2021tunable}.

\begin{figure*}[t]
\includegraphics[width=0.8\paperwidth]{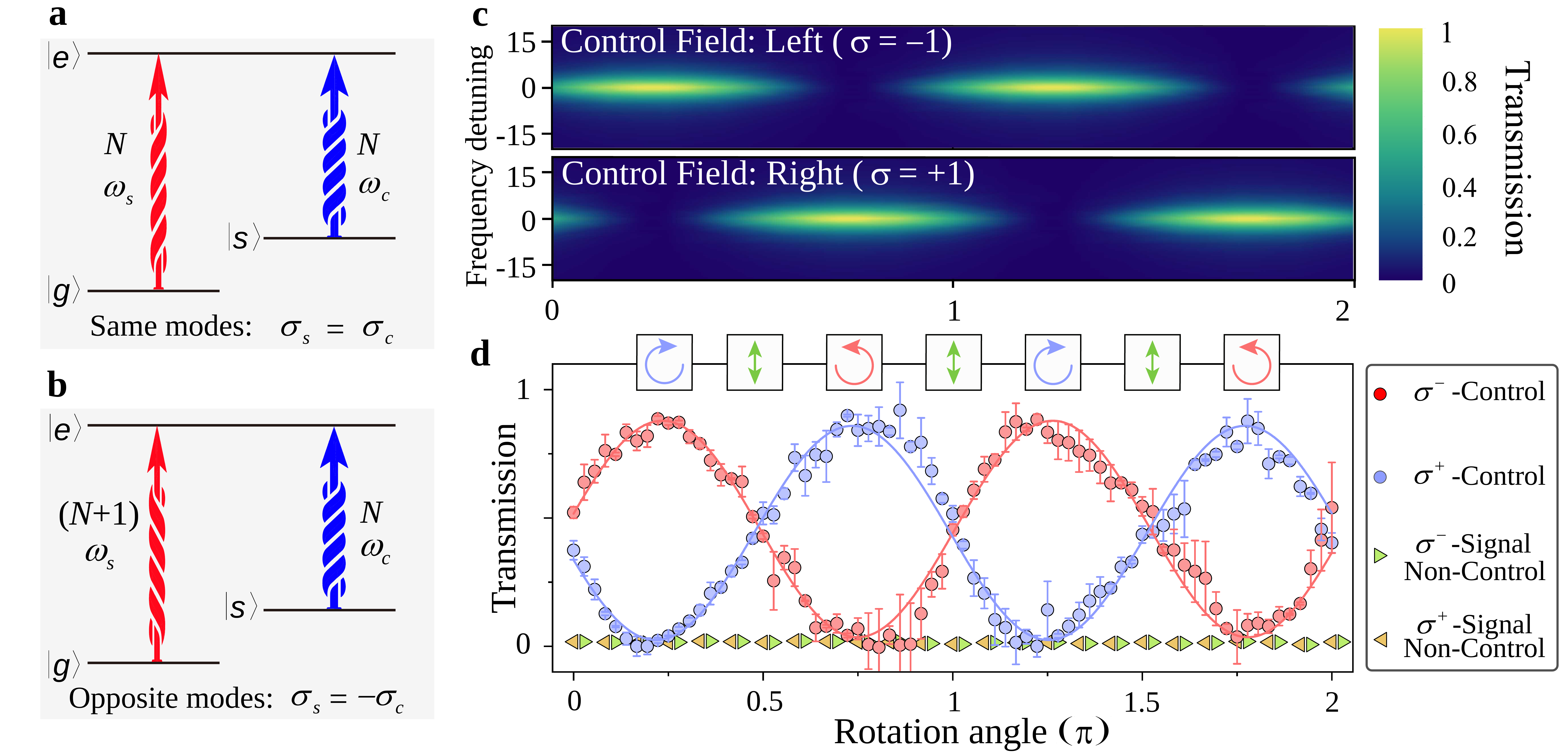}
\caption{\textbf{Experimental observation of the chirality transfer method.} 
\textbf{a,b}, Constructive and destructive interference pathways in this chirality-transfer approach. 
\textbf{c}, Simulation of the normalized chiral dichroism spectrum by scanning the signal field's chirality and frequency detuning. 
Here, $N$ denotes the number of control fields, $\hbar\omega_{s}=E_{e}-E_{g}$, and $\hbar\omega_{c}=E_{e}-E_{s}$.
The frequencies of the control and signal fields are near-resonance to the medium, and the control fields have enantiomeric polarized modes $\sigma=-1$ (the upper) and $\sigma=+1$ (the lower), respectively. 
\textbf{d}, Chirality-transfer circular-dichroism signal transmission spectrum $I_{z}^{(j)}(k)$ in the resonance region. 
Using a control field with a chosen helicity mode,  $\sigma^{-}$-polarized, the blue dots (red dots) represent the $\sigma^{+}$ ($\sigma^{-}$) polarized signal field. 
The green and yellow triangles are chiral dichroic transmission spectra of the enantiomeric signal field without control light. 
The error bars in the experimental data are estimated from Poisson statistics and represent $\pm$ one standard deviation. 
}
\label{Fig2}
\end{figure*}

\section*{Results }

\noindent\textbf{Chirality Transfer Mechanism} 

Helicity is a fundamental physical property of electromagnetic waves in isotropic dispersive media \cite{alpeggiani2018electromagnetic,bliokh2019topological}.
The chiral and topological properties of the medium can be characterized by helicity under the plane-wave approximation, and the involved light-matter interactions allow complex transformations in a dispersive medium \cite{fernandez2012helicity,fernandez2013electromagnetic,alpeggiani2018electromagnetic}.
Here, the helicity operator sets the general rules of chiral interaction in the process of chirality transfer.

In a neutral atomic system, two Maxwell equations for monochromatic light can be written as
\begin{equation}
\hat{\mathfrak{S}}\boldsymbol{\psi}=\omega\hat{\mathbf{M}}\partial_{t}\boldsymbol{\psi},\quad \hat{\mathbf{M}}=\left(
\begin{array}{cc}
\chi & i\mu \\
-i\varepsilon & \chi
\end{array}\right).
\label{eq:1}
\end{equation}
\noindent The helicity operator $\hat{\mathfrak{S}}=\hat{\mathbf{S}}\cdot\hat{\mathbf{p}}=\nabla \times$, and $\boldsymbol{\psi}=(\varepsilon\mathbf{E}, \mu\mathbf{H})^T$ is the wave function for the electromagnetic field. 
The operators $\hat{\mathbf{S}}$ and $\hat{\mathbf{p}}$ represent the spin and momentum operators, respectively.
The matrix $\hat{\mathbf{M}}$ is the modified constructive matrix,  where $\chi$ accounts for the \textit{chirality transfer} of the medium and $\varepsilon$ and $\mu$ are the relative permittivity and permeability of the medium, respectively.

If $\chi=0$ and $\varepsilon\propto\mu$, the helicity operator $\hat{\mathfrak{S}}$ provides a generator of the dual transformation \cite{bliokh2014magnetoelectric,alpeggiani2018electromagnetic}.
When considering the chirality-transfer-induced chiral response of the medium, the optical helicity is non-conserved in the two different eigenmodes.
Since the helical control field induces chirality transfer, the reconstructive matrix $\hat{\mathbf{M}}$ is non-Hermitian, exhibiting absorption in one eigenmode and gain in another.

The eigenmodes of the helicity operator in the medium can be expanded as a circularly polarized electric field $\mathbf{E}^{(j)}(\boldsymbol{r},t)=(1,i\sigma,0)\mathrm{exp}(i\boldsymbol{k}\cdot\boldsymbol{r}-i\omega t)$ and a magnetic field $\mathbf{H}^{(j)}(\boldsymbol{r},t)=-i\sigma Z^{-1}\mathbf{E}^{(j)}(\boldsymbol{r},t)$, where $k$ is the wave vector, $\boldsymbol{r}\equiv(x,y,z)$, $Z=\sqrt{\mu/\varepsilon}$, $j=L$ for $\sigma=-1$ and $j=R$ for $\sigma=+1$.
The mathematical structure of the incident field captures the winding of $\mathbf{E}(\boldsymbol{r},t)$ and $\mathbf{H}(\boldsymbol{r},t)$.
In our case, the rotating axis is along the $z$-direction.
The electric field of the paraxial wave beam becomes $\mathbf{E}(\boldsymbol{r},t)=\mathrm{cos}\theta~\mathbf{E}^{(L)}(\boldsymbol{r},t)+\mathrm{sin}\theta~\mathbf{E}^{(R)}(\boldsymbol{r},t)$, and the magnetic field has the same mathematical form \citep{supplimental}.
The $\mathrm{cos}\theta$ and $\mathrm{sin}\theta$ represent the projection of the plane wave to the different helical eigenstates. 

Figure \ref{Fig1} schematically illustrates the general chirality transfer mechanism in lossy and homogeneous media.
The atoms are initially populated in the ground state $\left|g,m_{F}\right\rangle$, where $\mathit{m_{F}}$ represents the magnetic quantum number of the atoms.
In Fig.~\ref{Fig1}a, the duality symmetry of the isotropic lossy medium breaks, with the optical chirality being conserved in this situation. 
Still, the medium will not provide a chiral response \cite{alpeggiani2018electromagnetic,bliokh2014magnetoelectric,bliokh2019topological}.
In the presence of the chiral ancillary field, as shown in Fig.~\ref{Fig1}b, the helical absorption signal measured after passing through the atomic medium is given by \citep{supplimental}
\begin{equation}
    I_{z}^{(j)}(\omega_s)=\mathrm{exp}\left[-2d|k_s|{\rm Im}\left[\left(1+\chi_{z}^{(j)}/2\right)\right]\Tilde{L}\right],
\label{eq:2}
\end{equation}
where $d$ represents the optical depth, $\omega_s$ is the optical frequency of the signal field, $\chi_{z}^{(j)}$ denotes the chirality transfer process, $L$ is the length of the medium, and $k_s=\omega_s/c$.
The control field introduces the chirality transfer with a specific non-vanishing helicity eigenvalue, exhibiting the enantio-sensitive absorption spectrum.
As shown in Fig.~\ref{Fig1}c, the atomic medium induced the helicity eigenmode with different eigenvalues respectively (left- and right-handed mode), and the chiral field alters the helicity and topological property of a homogeneous medium, leading to enantio-selective chiral dichroism.

By treating the chiral response as a perturbation, with time-dependent perturbation theory, we obtain the chiral susceptibility tensor for the medium in the sum-over-states form as $\chi_{z}^{(R)}$ and $\chi_{z}^{(L)}$ \citep{supplimental}.
The corresponding optical absorption experienced by the signal field is given by 
\begin{equation}
    \triangle\alpha_{z}\thickapprox d|k_{s}|{\rm Im}[(\Delta\chi_{z})], 
    \label{eq:3}
\end{equation}
where we define $\Delta\chi_{z}=\chi_{z}^{(L)}-\chi_{z}^{(R)}$ \citep{supplimental}.
Due to the chirality transfer, the chiral atomic medium presents distinct absorption spectra in different helicity eigenmodes of the signal field, as depicted in Fig.~\ref{Fig1}d and e.
Within the shaded region, the modulation of the chiral control field ceases to affect the signal effective absorption coefficient when the helicity eigenvalues of both fields are identical. 
Conversely, a significant variation in absorption versus frequency detuning occurs when the control and signal fields exhibit opposite helicity. 
The absorption spectrum of the enantiomeric signal field is suppressed by $\propto1/\triangle^{2}$, with the signal frequency detuning $\triangle$.

\begin{figure*}[t]
\includegraphics[width=0.8\paperwidth]{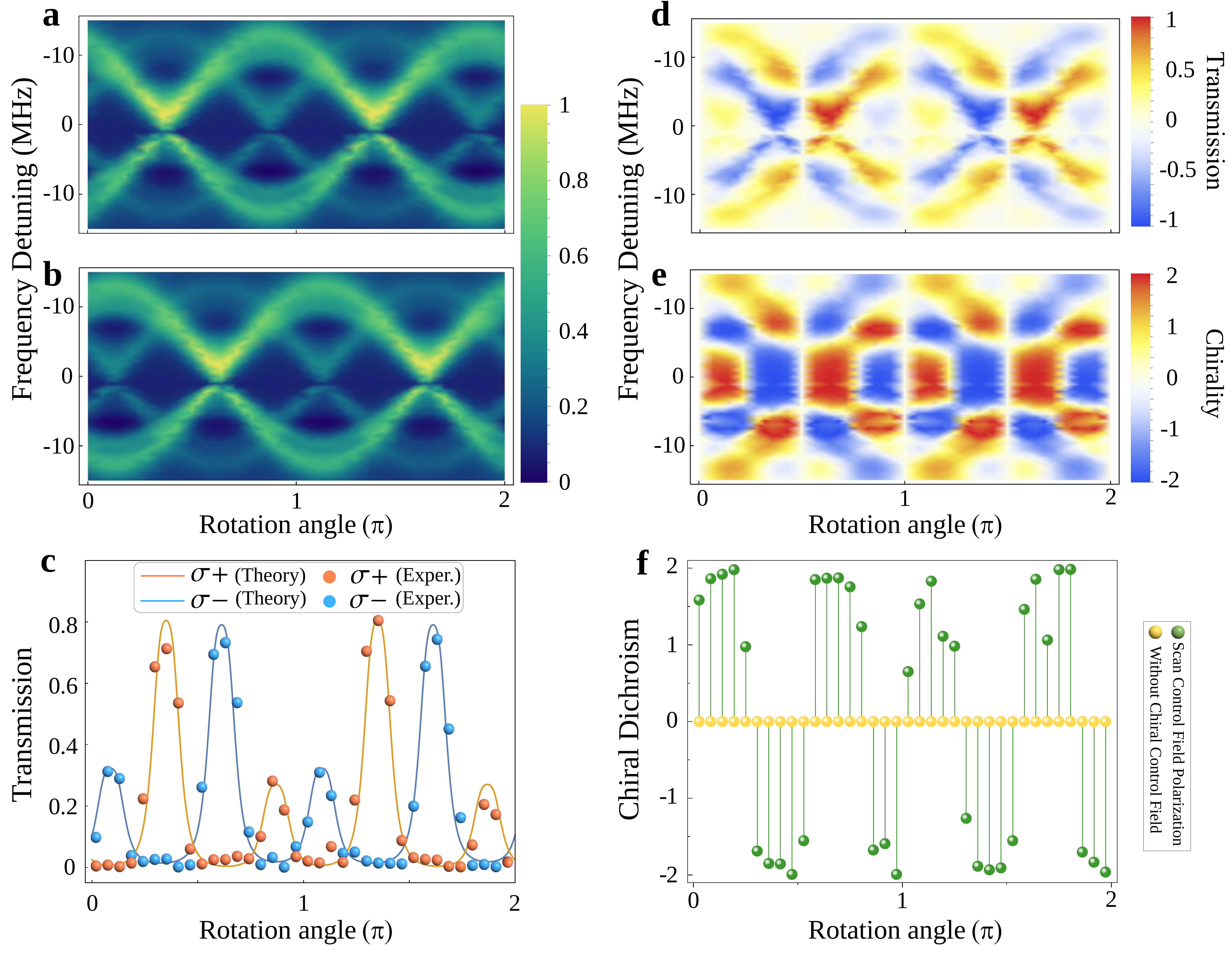}
\caption{\textbf{All-optical tunable control of the medium chirality.}
\textbf{a,b}, The absorption spectrum of the enantiomeric signal field, theoretically, versus the enantiomeric composition and frequency detuning of the control field.
\textbf{c}, Chiral dichroism transmission spectrum of the enantiomeric signal at the two-photon resonance detuning of $\Delta\omega_{s}=0$. The solid curves are theoretical fits by using Eqs.~(\ref{eq:2}, \ref{eq:4}). 
\textbf{d}, We simulate the absorption spectrum of the signal field with Eq.~(\ref{eq:3}).
\textbf{e}, The chiral coefficient of the atomic medium is calculated by Eq.~(\ref{eq:4}). 
\textbf{f}, Using Eq.~(\ref{eq:4}), we investigate the chiral dichroism spectrum of the medium, and its chirality achieves tunable modulation, which is consistent with our theoretical prediction.}
\label{Fig3}
\end{figure*}

The enantiomeric asymmetry of the medium results from the breaking of the spatial symmetry $\mathcal{P}$ induced by the control field.
Note that the conventional chiral field is solely used as the detection reagent in circular dichroism (CD) spectra applications \citep{ye2021enhancing,miles2021tools,ayuso2019synthetic,greenfield2006using}.
Our approach is based on a fully tunable chirality transfer mechanism with chiroptical interaction, which can control the medium chirality. 
This is distinct from the previous chiral state transfer methods with magnetic field-assisted Zeeman energy splitting, where the magnetic field induces an energy shift to break the spatial symmetry \citep{sayrin2015nanophotonic,scheucher2016quantum}. 
Note that the population condition illustrated in Fig.~\ref{Fig1} satisfies the Boltzmann distribution under thermal equilibrium; this is a common configuration in a chiral medium, and the additional magnetic field is not strictly required for breaking the $\mathcal{P}$-symmetry of the atomic population state. 

\noindent\textbf{Deterministic Chirality Transfer Process} 

Figures \ref{Fig2}a,b show the gain and loss interference pathways involved in the chirality transfer process. By tuning the helicity of the control field, we can achieve perfect constructive and destructive interference of the two paths and fully suppress or maximally enhance the signal absorption (Fig.~\ref{Fig1}d, e) in a selected enantiomer. Thus, we now have an effective chirality transfer process by dynamically controlling the optical helicity.

To quantify the chirality transfer, we observe the enantio-sensitive-dependent absorption spectra under the control field with specific helicity eigenmodes. The simulated CD spectrum of the signal field is shown in Fig.~\ref{Fig2}c, where the control and signal fields are near-resonance to the medium. The top row represents the control fields with a chirality of $\sigma=-1$, while the bottom row corresponds to a chirality of $\sigma=+1$.
When scanning the polarization $\theta$ of the signal field, we predict a periodic enantio-selective CD spectrum $I_{z}^{(j)}(\omega_s)$.
Furthermore, we find that the relative frequency detunings between the control and probe light affect the chiral response of the enantiomeric signal light. 
The atomic medium has the largest chiral response along the direction of light propagation only at the resonance point. 

Figure \ref{Fig2}d depicts the chirality-transferred CD signal transmission $I_{z}^{(j)}(\omega_s)$ versus the signal chirality parameter $\theta$ within the resonance area. 
Without the chiral control field, the atomic medium has a polarization-independent absorption response (corresponding to the green and yellow triangles).
The helicity of the medium is in a superposition mode of different eigenmodes, and the chirality is equal to 0 (achiral lossy medium).
When the helical control fields with opposite signs of helicity induce the chirality transfer, the signal field has opposite chiral responses at the chiroptical region (corresponding to the blue and red circles).
During the one-mode control, the left-handed ($\theta=\pi/4$) and right-handed ($\theta=3\pi/4$) circular-polarized signal fields exhibit the maximum and minimum absorption, and the period of variation is $\pi$.
The chiral transfer model, as shown in Fig.~\ref{Fig1}b, can be understood more clearly with a simplified model, describing the interference of two transfer pathways (as shown in Fig.~\ref{Fig2}a,b).
In addition, we can produce constructive or destructive interference of the two paths (Fig.~\ref{Fig2}a,b) and entirely suppress or maximally enhance the signal absorption in a selected enantiomer.
In the chiral pathway, the signal field is transparent due to the constructive interference of $N$ signal paths and $N$ control paths with the same helicity mode; the signal photons show the chiral gain; when the helicity modes have opposite signs, the destructive interference causes signals to be absorbed.
For $\theta=\pi/4$ or $\theta=3\pi/4$, the maximum CD spectrum difference is shown in Fig.~\ref{Fig2}d.
Meanwhile, the sign of the control field helicity flip induces the atomic medium into a different helicity mode and exhibits the opposite enantio-sensitive CD spectrum.

\noindent\textbf{Tunable control of the chirality transfer $\chi$}

By varying $\theta_{c}$ (polarization) and the strength of the relative control field, we can tunably control the chirality transfer process.
The enantiomeric signal field probes the chirality of medium, and we obtain the enantio-sensitive absorption spectrum of the medium enantiomers.
In Fig.~\ref{Fig3}a,b, the absorption spectrum of the enantiomeric field $\alpha_{z}^{(j)}$ ($j={R},\,{L}$) is determined by two parameters: the control field polarized angle $\theta_{c}$ and the frequency detuning $\bigtriangleup\omega$. 
The atomic chirality transfer effect is more pronounced for lower $\bigtriangleup\omega$, and the enantio-sensitive absorption spectrum is sensitive to the control field polarization angle $\theta_{c}$. 

As shown in Fig.~\ref{Fig3}c, we use the enantiomeric probe fields with opposite helicity mode to detect the $\alpha_{z}^{(j)}$ of the medium by scanning the $\theta_{c}$ of the control field.
As the direct observable measurement of the medium helicity, the enantiomeric signal field intuitively gives the chiral characteristics of the medium \cite{alpeggiani2018electromagnetic,bliokh2019topological}.
When $\theta=\pi/4$ and $\theta=3\pi/4$, the control fields are linearly polarized, and the corresponding chirality transfer efficiency becomes zero.

When the control field is in the different circularly polarized modes.
Their chirality transfer process yields different phase shifts $\phi^{L}\thickapprox\mathrm{Re}[\chi_{z}^{(L)}]$ of the signal field, which provides variations of the chiral response for the different modes \cite{alpeggiani2018electromagnetic}.
In this case, the enantiomeric polarized signal field have the maximum phase shift $\Delta\phi\propto\mathrm{Re}[\Delta\chi_{z}]=\pi/4$.
Thus, tuning the control field with period $\pi$, the enantio-sensitive absorption has four chiral regions.
Due to phase mismatch, the chiral light interaction between the upper ($-\pi/2-\pi/2$) and lower ($\pi/2-3\pi/2$) parts of the Poincaré sphere is different, resulting in enantiomeric absorption spectra with different amplitudes.
Furthermore, using Eq.~(\ref{eq:3}), we explore the relations between the circular dichroism absorption spectrum $\triangle\alpha_{z}$, polarization angle $\theta_{c}$ and frequency detuning $\bigtriangleup\omega$, as shown in Fig.~\ref{Fig3}d. 
The chiral absorption coefficient $\triangle\alpha_{z}$ is close to unity near the resonant region. 
Under the illumination of a particular enantiomeric control field, the medium can be fully polarized in one helicity mode and show a specific enantiomer with chiral symmetry. 
Figure \ref{Fig3}e illustrates the control-field-induced circular dichroism coefficient, where we define the effective chiral dichroism coefficient as 
\begin{equation}
C=2\frac{I_{z}^{(R)}(\omega_s)-I_{z}^{(L)}(\omega_s)}{I_{z}^{(R)}(\omega_s)+I_{z}^{(L)}(\omega_s)}.
\label{eq:4}
\end{equation}
The chiral dichroism of the medium presents a chiroptical response in the near-resonance region, while the chiroptical effect is weaker in the far-detuned region. 

By combining the results of Fig.~\ref{Fig3}c and Eq.~(\ref{eq:4}), we observe the tunable properties of the chiral medium by modifying the control field polarization. 
The experimental data in Fig.~\ref{Fig3}f are obtained under optical resonance.
The chiral dichroism parameter periodically flips with the change of the control field chirality, which makes our chiral manipulation tunable and also conforms to the theoretical prediction in Fig.~\ref{Fig3}e. 
Thus, we provide a method for arbitrarily manipulating the chirality of the medium to change the chiral enantiomer of the medium.

\begin{figure*}[t]
\includegraphics[width=0.8\paperwidth]{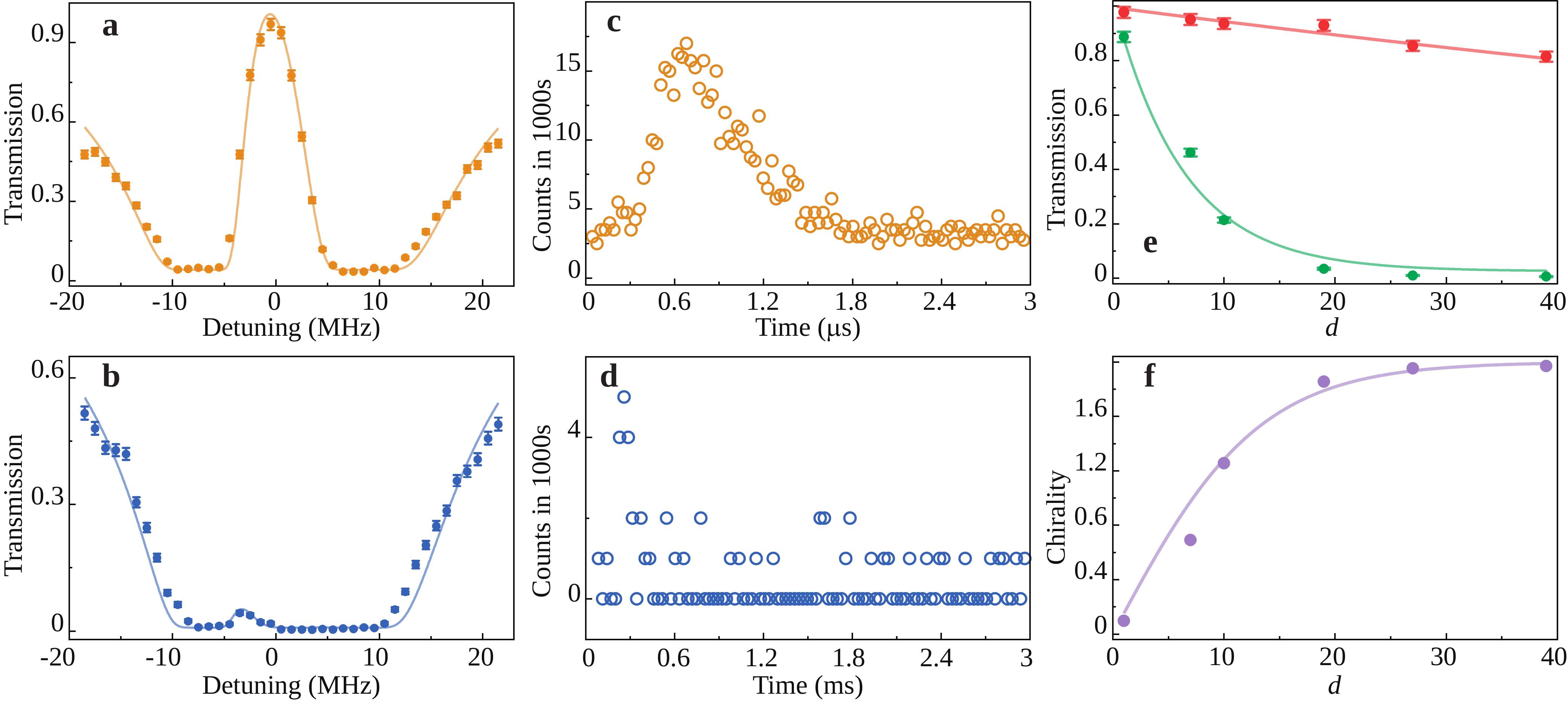}
\caption{\textbf{Dynamical properties of the chiral medium.}
We use a $\sigma^{+}$ circularly polarized field to control the chirality transfer process in the medium. 
\textbf{a},\textbf{b}, Single-photon enantio-sensitive transmission spectra of the signal field.
\textbf{a} and \textbf{b} correspond to the transition $\left|{g}\right\rangle \to\left|{e}\right\rangle $ in the $\sigma^{+}$ and $\sigma^{-}$ polarization cases, respectively.
\textbf{c,d}, Measurement of the second-order cross-correlation functions in the $\sigma^{+}$ (red dot) and $\sigma^{-}$ (blue dot) polarization cases.
\textbf{e}, The enantio-sensitive transmission spectrum of the $\sigma^{+}$ (red) and $\sigma^{-}$ (blue) signal field as a function of the optical depth at $\Delta\omega_{s}=0$. 
\textbf{f}, The overall chiral coefficient of the medium. 
The solid curves in (a-f) are theoretical fits by using Eqs.~(\ref{eq:3}, \ref{eq:4}), which agree well with the experimental data. 
The error bars in the experimental data are estimated from Poisson statistics and represent $\pm$ one standard deviation. }
\label{Fig4}
\end{figure*}

\noindent\textbf{Dynamical properties of the chiral medium} 

Figures~\ref{Fig4}a and~\ref{Fig4}b show the single-photon enantio-sensitive absorption spectrum of the signal field by scanning the detuning ${\Delta\omega_{s}}$, which corresponds to the same and opposite signal enantiomers, respectively.
The detuning ${\Delta\omega_{s}}$ is scanned from $-2{\rm \pi}\times18$ MHz to $\sim+2{\rm \pi}\times22$ MHz, with a frequency resolution of $2{\rm \pi}\times1$ MHz. 
In the vicinity of ${\Delta\omega_{s}}\sim0$, we observe the transmission spectrum for the $\sigma^{+}$ and $\sigma^{-}$ polarization signal field cases, respectively (which was theoretically predicted in Fig.~\ref{Fig1}d). 
Therefore, a high chirality transfer rate is available when the single-signal photons are resonant with the transition $\left|{g}\right\rangle \to\left|{e}\right\rangle$.
In the two-photon resonance region, the transmitted single-signal photon pulse is characterized by the measured second-order cross-correlation function $g^{(2)}(\tau)$, where $\tau$ is the time delay. 
Keeping the chiral configuration of the experimental scheme unchanged, we obtain the results in Fig.~\ref{Fig4}c,d. 
By performing the Hanbury-Brown--Twiss experiment, we confirm that our chirality transfer approach can maintain the single-photon nature of the signal field. 
We obtain a heralded auto-correlation parameter $\ensuremath{\alpha}$ of $0.0201\pm0.0008$ for the input heralded single-photon source and $0.0232\pm0.0011$ for a single photon passing through the medium along the $\hat{z}$-direction \citep{supplimental}.

According to Eq.~(\ref{eq:3}), the chiral dichroism spectrum of the signal field is dominated by the imaginary part of the susceptibility of different enantiomers. 
For a circularly polarized signal field, the susceptibility of the medium is affected by the optical depth $d$ of the medium \citep{supplimental}, and the chiral absorption spectrum is proportional to $\mathrm{exp}(-d)$. 
Here, $d$ is a dimensionless real number determined by the atomic density and the medium length in the $\hat{z}$-direction.
While keeping the experimental setup unchanged in the near-resonance region, we simulate the effect of changing the medium density on the absorption of the signal light enantiomers. 
The transmitted spectrum through the chiral medium is shown in Fig.~\ref{Fig4}e. 
Considering the enantiomeric sensitive absorption spectrum and chiral coefficient in Eq.~(\ref{eq:4}), we observe the overall chiral coefficient of the medium in Fig.~\ref{Fig4}f. 
Note that when 0 $<$ $d$ $<$ 20, the medium exhibits an asymmetric absorption of chiral enantiomers.
This reveals that our chiral transfer method is feasible and independent of the medium. 
When $d$ $>$ 20, the medium exhibits the largest enantiomeric absorption difference and is completely chiral polarized to the helicity eigenmode. 
When the $d$ increases, the medium chirality remains unchanged but introduces additional signal loss. 
\section*{Discussion}
We proposed and experimentally realized a new optical media chirality transfer and manipulation approach. 
These results lay the foundation for controlling asymmetry synthesis and open a new door for enantio-sensitive manipulation in atomic media.
This chiral mechanism could be applied to dual particles (e.g., molecule and atom), and is not limited by the number of particles. 
This includes the selective detection of enantiomers with specific rotation in non-enantiomeric samples \citep{eibenberger2017enantiomer,perez2017coherent}.
The proposed chirality transfer method offers new opportunities for the enantiomeric selected excitation. 
This transfer and manipulation mechanism can also be used to imprint chirality efficiently on achiral matter \citep{ordonez2019propensity}, overcoming the limitation of the recent proposal of molecular rotations to other degrees of freedom \citep{owens2018climbing}. 
Our chirality transfer approach also provides an opportunity to transfer an achiral medium to a chiral medium, overcoming the limitation of rotational degrees of freedom to other degrees of freedom. 
More extensively, this chiral transfer mechanism opens the prospect of studying laser-driven achiral-chiral phase transitions.

\section*{Method}
\noindent\textbf{Atomic energy level Configuration}

\noindent Here, we derive the field-theoretical form of the chirality transfer method for inhomogeneous media. 
In our theoretical model, the signal field passes through an ensemble of cold atoms illuminated by the control field. 
In Fig.~\ref{Fig1}a,b,  $\left|g\right\rangle=\left|5^{2}S_{1/2},F=2\right\rangle$,  $\left|s\right\rangle=\left|5^{2}S_{1/2},F=3\right\rangle$, and $\left|e\right\rangle=\left|5^{2}P_{1/2},F=3\right\rangle$.
Note that the interaction cross section between the chiral field and medium depends on the magnetic quantum number of the specific energy level, but not on the respective Land\'{e} factor \citep{sayrin2015nanophotonic}.
The inducing control field with $\sigma^{+}$-polarization thus builds an effective transfer channel $$\left|s,m_{F}=i\right\rangle \leftrightarrow\left|e,m_{F^{\prime}}=i+1\right\rangle $$ line in Fig.~\ref{Fig1}b. 

For an input signal near-resonance to the energy level $|g\rangle \rightarrow|e\rangle$ and $F_{g}<F_{e}$, the chiral medium builds an effective transfer channel, in which the signal chiral character is directed by the transfer $$\left|g,m_{F}=i\right\rangle \leftrightarrow\left|e,m_{F^{\prime}}=i-1\right\rangle $$ or $$\left|g,m_{F}=i\right\rangle \leftrightarrow\left|e,m_{F^{\prime}}=i+1\right\rangle $$.
Due to the interference of the transition paths between the two states ($|g\rangle $ and $|s\rangle $) and the excited state $|e\rangle $, the medium shows transparent properties to the $\sigma^{+}$-polarized signal. 
In contrast, the chiral medium absorbs the opposite chiral polarized signal field (with $\sigma^{-}$-transition), resulting from the forbidden ground states interference effect. 
The medium exhibits the opposite chiral property when the chiral control field has opposite helicity.

\noindent\textbf{The enantio-selective chiral response of the atomic medium}

\noindent The signal transmission spectrum has the form $$T_{z}^{(j)}=\mathrm{exp}(-2d|k_{s}|\mathrm{Im}[(1+\chi^{(j)}_{z}/2)]\Tilde{L}),$$ and the chiral susceptibilities in different enantiomers can be described by the following relations:
\begin{equation}
    \chi^{(L)}_{z}=\chi^{(L)},
\end{equation}
\begin{equation}
\chi^{(R)}_{z}=\chi^{(R)}{-\frac{N\left|\mu_{e_{-3},g_{-2}}\right|^{2}}{5\hbar\varepsilon_{0}(\Delta\omega_{s}+i\gamma_{ge})}}\;,
\end{equation}
\\where the $\chi^{(j)}$-components are
\begin{equation}
  \chi^{(R)}=\frac{4N}{5\hbar\varepsilon_{0}}\sum\limits _{i=-2}^{1}\frac{\left|\mu_{e_{i+1},g_{i}}\right|^{2}(\delta+i\gamma_{gs})}{\left|\Omega_{c,i}\right|^{2}-4(\delta+i\gamma_{gs})(\Delta\omega_{s}+i\gamma_{ge})},
\end{equation}
and
\begin{equation}
\chi^{(L)}=\frac{4N}{5\hbar\varepsilon_{0}}\sum\limits _{i=-1}^{2}\frac{\left|\mu_{e_{-i-1},g_{-i}}\right|^{2}(\delta+i\gamma_{gs})}{\left|\Omega_{c,i-2}\right|^{2}-4(\delta+i\gamma_{gs})(\Delta\omega_{s}+i\gamma_{ge})}.
\end{equation}
Here, $\mu_{e_{i+1},g_{i}}$ and $\mu_{e_{-i-1},g_{-i}}$ are the corresponding dipole moments for the transitions $$\left|g,m_{F}=i\right\rangle \to\left|e,m_{F^{\prime}}=i+1\right\rangle$$ and $$\left|g,m_{F}=-i\right\rangle \to\left|e,m_{F^{\prime}}=-i-1\right\rangle, $$
where $\Omega_{c, i}$ denotes the Rabi frequency of the control field driving the transition $$\left|s,m_{F}=i\right\rangle \to\left|e,m_{F^{\prime}}=i+1\right\rangle,$$ with $i\in\{-3,2\}$ for the $\sigma^{+}$- and $\sigma^{-}$-propagation cases. 
Also, $N$ is the atomic density, $\gamma_{gs}$ ($\gamma_{ge}$) is the mean dephasing rate between levels $\left|g\right\rangle $ and $\left|s\right\rangle $ ($\left|g\right\rangle $ and $\left|e\right\rangle $). 
In the absence of a magnetic field, the detunings between the coupling and signal field and the corresponding transition become $$\Delta\omega_{c}=\omega_{c}-\omega_{se}$$ and  $$\Delta\omega_{s}=\omega_{s}-\omega_{ge},$$ yielding a two-photon detuning $\delta=\Delta\omega_{s}-\Delta\omega_{c}$. 
The contributions of the $\chi_{z}^{(j)}$ part are nearly identical for both cases. 

We then numerically simulate the chiral signal field that passes through the ensemble. 
The signal-field amplitude launched into the atoms is assumed to be constant in time, thus driving the ensemble of atoms with fixed Rabi frequency. 
The thermal motion of atoms and the stray magnetic field can be neglected in the cold atomic ensemble.

\noindent\textbf{Experiment setup and polarization calibration}

\noindent Our all-optical enantio-selectivity and chirality transfer approach is validated by optically regulating cold Rubidium atoms. 
The atoms are loaded from free atomic gas into a magneto-optical trap via a laser cooling strategy \citep{ye2022long,zhang2017quantum,ding2013single,dong2020temporal}. 
Two external cavity diode laser sources (Toptica, DL pro 795 nm) are used as the control and signal fields, corresponding to the D1 transitions of the Rb atoms. 
Both lasers are coupled to the system through fibers, and a small angle (less than $3^{\circ}$) exists between the control and signal fields. 
The local chirality of the enantiomeric medium is controlled by the helicity of the control field in a cold atomic ensemble, and the chirality of the medium is verified by changing the helicity of the signal single photon. 

The $\sigma^{+}$ polarized signal photon is converted to the $\sigma^{-}$ polarized photon used by a half and quarter waveplate.
Both signal and control fields are propagated along the $\hat{z}$-direction.
Initially, we set the linear polarization along the $\hat{x}$-axis, representing the horizontal polarization.
Here $\theta$ is the angle between the electric field of the light beam, and the fast axis of the quarter-wave plate is parallel along the $\hat{x}$-axis.
Then, varying $\theta$ from 0 to $2\pi$ shows that the incident light polarization travels a closed path on the Poincaré sphere.

\noindent\textbf{Acknowledgments}

\noindent We thank Prof. Wei Yi, Prof. Chun-Hua Dong, Dr. Wei Zhang, Dr. Meng-Jun Hu, Dr. Ying-Hao Ye, and Dr. Lei Zeng for fruitful discussions. 

\noindent\textbf{Funding} 

\noindent M.X.D. acknowledges funding from the National Natural Science Foundation of China (12204461).

\noindent D.S.D. acknowledges funding from the National Key Research and Development Program of China (2022YFA1404002), the National Natural Science Foundation of China (Grant No. U20A20218), the Major Science and Technology Projects in Anhui Province (Grant No. 202203a13010001), and the Youth Innovation Promotion Association of Chinese Academy of Sciences under Grant No. 2018490. 

\noindent B.S.S. acknowledges funding from the National Natural Science Foundation of China (Grant No. 11934013), the Innovation Program for Quantum Science and Technology (2021ZD0301100), and Anhui Initiative in Quantum Information Technologies (AHY020200).

\noindent F.N. is supported in part by Nippon Telegraph and Telephone Corporation (NTT) Research, the Japan Science and Technology Agency (JST) [via the Quantum Leap Flagship Program (Q-LEAP), and the Moonshot R\&D Grant Number JPMJMS2061], the Asian Office of Aerospace Research and Development (AOARD) (via Grant No. FA2386-20-1-4069), and the office for Naval Research (ONR) Global (via Grant No. N62909-23-1-2074).

\noindent\textbf{Data availability}

\noindent All data needed to evaluate the conclusions in the paper are presented in the paper. 
Additional data related to this paper may be requested from the authors.

\noindent\textbf{Author contributions} 

\noindent D.S.D., M.X.D., and E.Z.L. conceived the idea with discussions with B.S.S.. E.Z.L. and M.X.D. carried out the experiments with assistance from W.H.Z..
All authors contributed to the discussions and analysis of the results. 
E.Z.L. and M.X.D. wrote the manuscript with contributions from F.N., D.S.D., and B.S.S.. 
M.X.D., D.S.D., B.S.S., and F.N. supervised the project.

\noindent\textbf{Competing interests}

\noindent The authors declare no competing interests.

\bibliographystyle{apsrev4-2}

%
\end{document}